\documentclass[twocolumn,showpacs,unsortedaddress,showkeys,preprintnumbers,amsmath,amssymb]{revtex4}
\usepackage{graphicx}
\usepackage{dcolumn}
\usepackage{bm}

\begin{document}

\title{Inference of a nonlinear stochastic model of the cardiorespiratory interaction}
\author{V.N. Smelyanskiy$^1$}
\author{D.G. Luchinsky$^2$}
\author{A. Stefanovska$^{2,3}$}
\author{P.V.E. McClintock$^2$}

\affiliation{$^1$NASA Ames Research Center, Mail Stop 269-2,
Moffett Field, CA 94035, USA}

\affiliation{$^2$Department of Physics, Lancaster University,
Lancaster LA1 4YB, UK}

\affiliation{$^3$Faculty of Electrical Engineering, University of
Ljubljana, Tr\v{z}a\v{s}ka 25, 1000 Ljubljana, Slovenia}

\date{\today}

\begin{abstract}
A new technique is introduced to reconstruct a nonlinear
stochastic model of the cardiorespiratory interaction. Its
inferential framework uses a set of polynomial basis functions
representing the nonlinear force governing the system
oscillations. The strength and direction of coupling, and the
noise intensity are simultaneously inferred from a univariate
blood pressure signal, monitored in a clinical environment. The
technique does not require extensive global optimization and it is
applicable to a wide range of complex dynamical systems subject to
noise.
\end{abstract}


\maketitle

Heart rate variability (HRV) is an important dynamical phenomenon
in physiology. Altered HRV is associated with a range of
cardiovascular diseases and increased mortality \cite{Camm:96},
and its parameters are starting to be used as a basis for
diagnostic tests. However, signals acquired from the human
cardiovascular system (CVS), being derived from a living organism,
arise through the interaction of many dynamical degrees of freedom
and processes with different time scales
\cite{Winfree:80Glass:88}. Thus HRV is attributable to the mutual
interaction of a large number of oscillatory processes. Among
them, the effect of respiration on heart rate has been the most
intensively studied. The physiological mechanisms have recently
been reviewed \cite{Eckberg:03} and include e.g.\ modulation of
the cardiac filling pressure as a result of changes of
intrathoracic pressure during respiratory
movements~\cite{Visscher:24}, direct respiratory ordering of
autonomic outflow~\cite{Eckberg:03}, and baroreceptor feedback
control~\cite{deBoer:87}.

An important feature of these processes is that they are
nonlinear, time-varying, and subject to fluctuations
~\cite{Saul:88a,Chon:96Suder:98,Stefanovska:99a}. For such systems
deterministic techniques fail to yield accurate parameter
estimates~\cite{Kostelich:92McSharry:99a}. Additionally, models of
the cardiovascular interactions are not usually known exactly from
first principles and one is faced with a rather broad range of
possible parametric models to consider
~\cite{deBoer:87,Clynes:60Baselli:88TenVoorde:95Seidel:95Cavalcanti:96Kotani:02}.
Inverse approaches, in which dynamical properties are analysed
from measured data have recently been considered. A variety of
numerical techniques have been introduced to analyse
cardio-respiratory interactions using e.g.\ linear
approximations~\cite{Berger:89Taylor:01Mukkamala:01Chon:01b},
estimations of either the strength of some of the nonlinear
terms~\cite{Jamsek:03Jamsek:04}, the occurrence of
cardio-respiratory synchronization~\cite{Schaefer:98Janson:01} or
the directionality of coupling~\cite{Rosenblum:02Palus:03a}.
Hitherto, modelling approaches have not been used interactively in
conjunction with time series analysis methods. Rather, the latter
have each focussed on a particular dynamical property, e.g.\
synchronization, or nonlinearities, or directionality.

In this Letter we introduce an approach to the problem that
combines mathematical modelling of system dynamics and extraction
of model parameters directly from measured time series. In this
way we estimate simultaneously the strength, directionality of
coupling and noise intensity in the cardio-respiratory
interaction. The technique reconstructs the nonlinear system
dynamics in the presence of fluctuations. In addition, the method
provides optimal compensation of dynamical noise-induced errors
for continuous systems while avoiding extensive numerical
optimization. We demonstrate the approach by using a univariate
blood pressure (BP) signal for reconstruction of a nonlinear
stochastic model of the cardio-respiratory interaction. The
results are verified by analysis of data synthesized from the
inferred model.

The problems faced in the analysis of CVS variability are common,
not only to all living systems, but also to all complex systems
subject to fluctuations, e.g.\ molecular motors \cite{Visscher:99}
or coupled matter--radiation systems in astrophysics
\cite{Christensen:02}. Yet there are no general methods for the
dynamical inference of stochastic nonlinear systems. Thus the
technique introduced in this paper will be of wide applicability.

\begin{figure}[h]   
   \begin{center}
 \includegraphics[width=8cm,height=5cm]{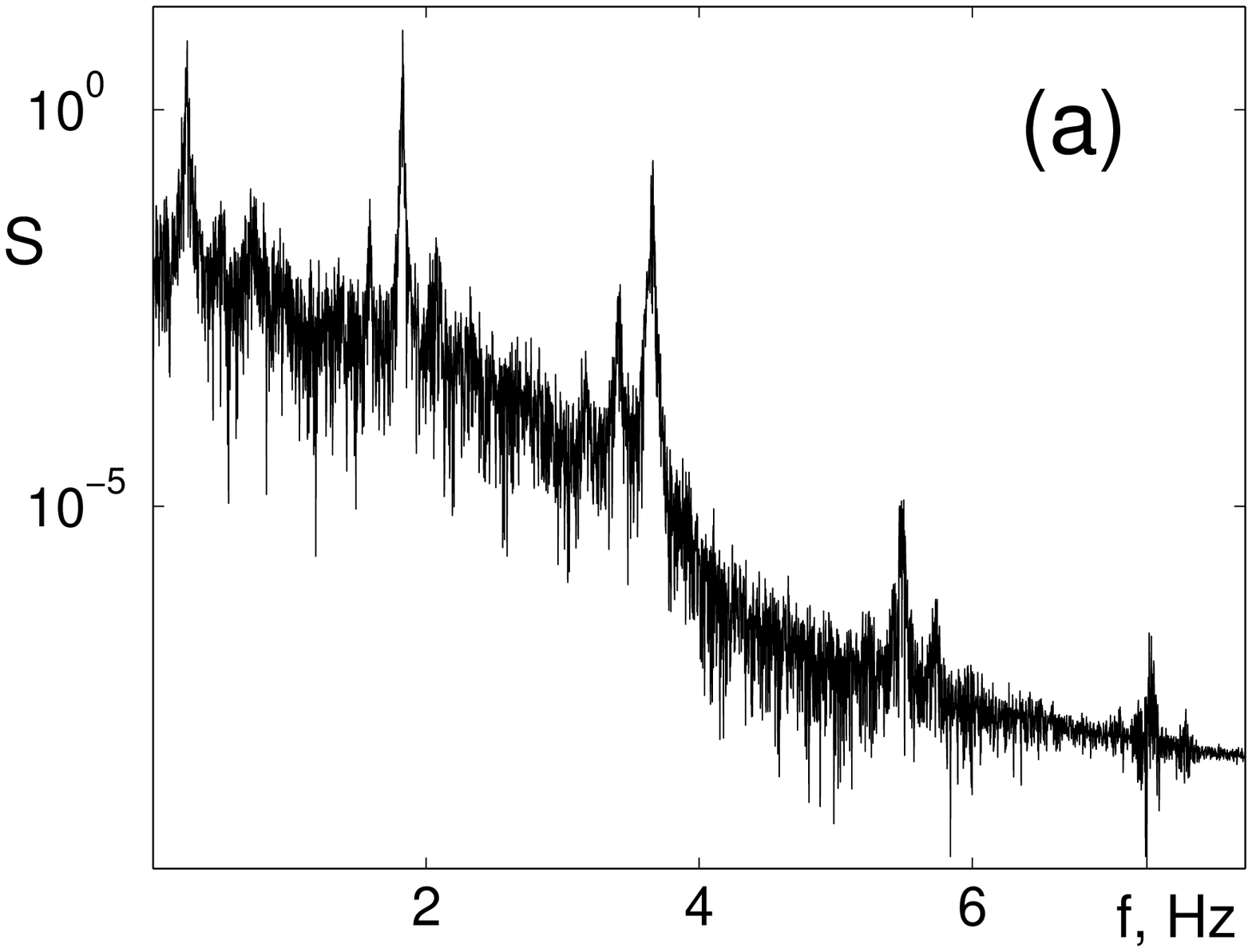}
 \includegraphics[width=7.5cm,height=4cm]{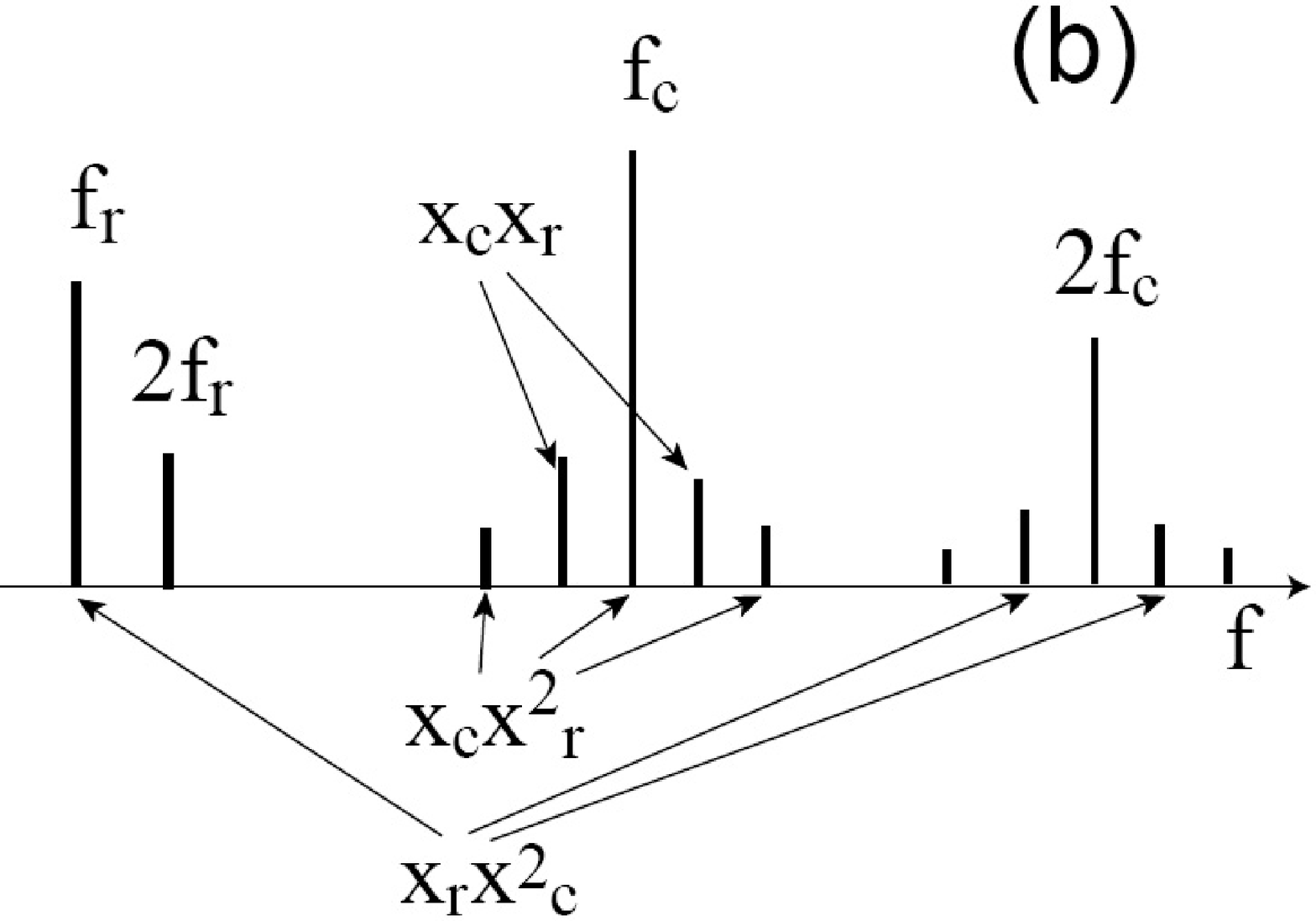}
   \end{center}
 \caption{\label{fig:summary} (a) Power spectrum of the venous
blood pressure (BP) data after filtration through Butterworth
filters: low-pass of the 4$^{th}$ order, with a cut-off frequency
of 3 Hz; and high-pass of the 2$^{nd}$ order with cut-off
frequency of 0.03 Hz. (b) Summary of the main combinatorial
frequencies of the cardiac and respiratory components observed in
the BP signal. The correspondence between the nonlinear
interaction terms of the model (\ref{eq:2vdp_1}) and the
frequencies observed in the time-series data are shown by arrows.}
 \end{figure}

We use public domain data to illustrate the idea. We analyse
central venous blood pressure data, record 24 of the MGH/MF
Waveform Database available at www.physionet.org. Its spectrum,
shown in Fig. \ref{fig:summary}(a), exhibits two basic frequencies
corresponding to the respiratory, $f_r\approx 0.2$ Hz, and
cardiac, $f_c\approx 1.7$ Hz, oscillations; the higher frequency
peaks are the 2nd, 3rd and 4th harmonics of the cardiac
oscillation. We note that the relative intensity and position of
these peaks vary from subject to subject, with the average
frequencies for healthy subjects at rest being around $0.2$ and
$1.1$ Hz for respiration and heart rate respectively.

We must bear in mind that CVS power spectra also contain lower
frequency
components~\cite{Camm:96,Stefanovska:97aTaylor:98Stefanovska:01b}.
In practice, parametric modelling is usually restricted to a
specific part of the power spectrum. Because our interest here
centres on the cardio-respiratory interaction, we select for study
the frequency range that includes the main harmonics of cardiac
and respiratory oscillations $f_c$ and $f_r$ and their
combinational frequencies as shown in Fig. \ref{fig:summary}(b).
In addition, we assume that the two higher basic frequency
components observed in all CVS signals
\cite{Stefanovska:99a,Stefanovska:01a} can be separated. Hence the
blood pressure signal can be considered in the first approximation
as a sum of the cardiac and respiratory oscillatory components
$s(t)=s_c(t)+s_r(t)$. Accordingly, we use a combination of
zero-phase forward and reverse digital filtering based on
Butterworth filters to decompose \cite{Decomposition} the blood
pressure signal into 2-dimensional time series $\lbrace {\bf
s}(t_k)=(s_c(t_k),s_r(t_k)),\,\,t_k=k h,\,k=0:K\rbrace$. The time
series represent the contributions of cardiac and respiratory
oscillations to the blood pressure on a discrete time grid. A
window consisting of 18000 points of the original signal, sampled
at 360 Hz, was resampled at 90 Hz. Hence the signal considered for
inference was of length 500~s, with a step size of $h=1/90$~sec.

Following the suggestion of coupled oscillators
\cite{Stefanovska:99a,Stefanovska:01a}, we now choose the simplest
model that can reproduce this type of oscillation: two nonlinearly
coupled systems with limit cycles on a plane
\begin{eqnarray}
  \label{eq:2vdp_1}\left\{%
\begin{array}{ll}
    \dot x_r = a_1x_r+y_r, & \dot y_r = \alpha_i\phi_i({\bf x},{\bf y}) + \sqrt{D_{1j}}\xi_j, \\
    \dot x_c = a_2x_c+y_c, & \dot y_c = \beta_i\phi_i({\bf x},{\bf y}) + \sqrt{D_{2j}}\xi_j
\end{array}%
\right.
\end{eqnarray}
are included. Here $\xi_j(t)$ are zero-mean white Gaussian noises,
and the summation is taken over repeated indexes $i=1,...,22$ and
$j=r,c$. The base functions are chosen in the form
\begin{eqnarray}
\label{eq:basis}
  &&\hspace{-0.5cm}\phi=\{1, x_r, x_c, y_r, y_c, x_r^2,
      x_c^2, y_r^2, y_c^2, x_ry_r, x_cy_c, x_r^3, \nonumber\\
 &&\hspace{-0.6cm} x_c^3, x_r^2y_r, x_c^2y_c, x_ry_r^2, x_cy_c^2, y_r^3,
 y_c^3, x_rx_c, x_r^2x_c, x_rx_c^2 \},
\end{eqnarray}
\noindent that includes nonlinear coupling terms up to 3rd order.
We assume that the measurement noise can be neglected. The two
dynamical variables of the model (\ref{eq:2vdp_1}), $x_r(t)$ and
$x_c(t)$ correspond to the two-dimensional time-series, ${\bf
s}(t)=\{s_r(t), s_c(t)\}$, introduced above. Using
(\ref{eq:2vdp_1}) the remaining two dynamical variables ${\bf
y}(t)=\{y_r(t),y_c(t)\}$ can be related to the observations
$\{{\bf s}(t_k)\}$ as follows
\begin{eqnarray}
 \label{eq:embedding}
 b_ny_n(t_k) = \frac{s_n(t_k+h)-s_n(t_k-h)}{2h} +a_{n}s_n(t_k),
\end{eqnarray}
where $n=r,c$. Parametric presentation (\ref{eq:2vdp_1}) with a
special form of embedding (\ref{eq:embedding}) allows one to infer
a wide class of dynamical models including e.g.\ the van der Pol
and FitzHugh-Nagumo models. Furthermore, it allows physiological
interpretation of the model parameters.

Using (\ref{eq:embedding}) we can reduce the original problem of
characterizing the cardio-respiratory interaction to that of
inferring the set of unknown parameters ${\cal M}=\{{\bf
c},{\bf\hat D}\}$ of the coupled stochastic nonlinear differential
equations
\begin{eqnarray}
 &&\dot {\bf y} = {\bf \hat U}({\bf s},{\bf y}){\bf c} + \sqrt{\bf\hat D}{\boldsymbol
 \xi}(t).
 \label{eq:cardio-respiratory}
\end{eqnarray}
\noindent Here $ {\boldsymbol \xi}(t)$ is a two-dimensional
Gaussian white noise with independent components mixed with
unknown correlation matrix ${\bf\hat D}$. The matrix ${\hat {\bf
U}}$ will have the following block structure
\begin{eqnarray}
 \label{eq:block_structure_1}
&&\hspace{-0.25in}{\hat {\bf U}}=\left[
  \left[\begin{array}{ll}
    1&0 \\
    0&1 \\
  \end{array}\right],
  \left[\begin{array}{ll}
    x_r&0 \\
    0&x_r \\
  \end{array}\right]   \ldots
  \left[\begin{array}{ll}
    x_rx_c^2&0 \\
    0&x_rx_c^2 \\
  \end{array}\right]
  \right].
\end{eqnarray}
The vector of unknown coefficients ${\bf
c}=\{\alpha_1,\beta_1,...,\alpha_{22},\beta_{22}\}$ has the length
$M=2B$, where $B=22$ diagonal blocks of size $2\times 2$ formed by
the basis functions (\ref{eq:basis}).

The model parameters can be obtained by use of our novel method of
dynamical inference of stochastic nonlinear models. The method is
based on the Bayesian technique. Details, and a comparison with
the results of earlier research, are given
elsewhere~\cite{Smelyanskiy:submitted}. Here we describe briefly
the main steps in applying the method to inference of
cardio-respiratory interactions. First, one has to define the
so-called {\it likelihood} function $\ell({\bf y}|{\cal M})$: the
probability density to observe the dynamical variables ${\bf
y}(t)$ under the condition that the underlying dynamical model
(\ref{eq:cardio-respiratory}) has a given set of parameters ${\cal
M}$. We suggest that, for a uniform sampling scheme and a
sufficiently small time step $h$, one can use results from
\cite{Graham:77} to write the logarithm of the likelihood function
as
\begin{eqnarray}
 && \hspace{-0.3in}-\frac{2}{K}\log  \ell({\bf y}|{\cal M})  =
\ln\det{\hat{\bf D}}
+\frac{h}{K}\sum_{k=0}^{K-1}\left[\,{\bf v}({\bf y}_k){\bf c}\right.\label{eq:likelihood} \\
  &&\hspace{-0.2in}\left. +(\dot{\bf y}_{k}  - {\hat {\bf U}}_k \, {\bf c})^T \, {\hat {\bf D}}^{-1} \,
    (\dot{\bf y}_{k}  - {\hat {\bf U}}_k \, {\bf c}))\right]+ N\ln(2\pi h).\nonumber
\end{eqnarray}
\noindent Here $ { \bf \hat U}_{k} \equiv {\bf \hat U}({\bf
y}_{k})$, $\dot {\bf y}_{k}\equiv h^{-1} ({\bf y}_{k+1}-{\bf
y}_k)$ and the vector ${\bf v}({\bf x})$ has components
\[\textrm{v}_{m}({\bf x})=\sum_{n=1}^{N}\frac{\partial U_{n\,m}({\bf
x})}{\partial x_n},\quad m=1:M.\]

\noindent Note that the form of (\ref{eq:likelihood}) differs from
the cost function in the method of least-squares: the term
involving ${\bf v}$ provides optimal compensation of noise-induced
errors ~\cite{Smelyanskiy:submitted}. In the next step one has to
summarize {\it a priori} expert knowledge about the model
parameters in the so-called {\em prior} PDF,
$p_{\textrm{pr}}({\cal M})$. We assume $p_{\textrm{pr}}({\cal M})$
to be Gaussian with respect to the elements of ${\bf c}$ and
uniform with respect to the elements of ${\bf \hat D}$.

Finally, one can use the measured time-series ${\bf y}$ to improve
the {\it a priori} estimation of the model parameters. The
improved knowledge is summarized in the {\em posterior}
conditional PDF $p_{\textrm{post}}({\cal M}|{\bf y})$, which is
related to the {\em prior} PDF via Bayes' theorem
\begin{equation}
    p_{\textrm{post}}({\cal M}|{\bf y}) = \frac{{\ell}({\bf y}|{\cal M}) \,
    p_{\textrm{pr}}({\cal M})}{\int \ell({\bf y}|{\cal
    M}) \, p_{\textrm{pr}}({\cal M}) \, {\rm d}{\cal M}}.
    \label{eq:Bayes}
\end{equation}
For a sufficiently large number of observations,
$p_{\textrm{post}}$ is sharply peaked at a certain most probable
model  $\cal M={\cal M}^{\ast}$, providing a solution to the
inference problem.

To find this solution we substitute the prior
$p_{\textrm{pr}}({\cal M})$ and the likelihood $\ell({\bf y}|{\cal
M})$ into (\ref{eq:Bayes}) and perform the optimization by
differentiation of the resulting expression with respect to ${\bf
\hat D}_{\textsf{y}}^{n n'}$ and $c_m$, yielding the final result
\begin{eqnarray}
 &&\hspace{-0.3in}{\bf \hat D}_{\textsf{post}}^{n n'}({\bf c})
 \equiv \frac{1}{K} \, \sum_{k=0}^{K-1} \left[ {\dot {\bf y}}_{k} -
 {\hat {\bf U}}_k \, {\bf c} \right]_n \left[ {\dot {\bf y}}_{k} -
  \, {\hat {\bf U}}_k{\bf c}
 \right]^{T}_{n'},\label{eq:updateD} \\
 &&\hspace{-0.3in}{\bf c}^{\prime}_{\textrm{post}}({\bf \hat D})={\hat {\boldsymbol
 \Xi}}^{-1}_\textsf{y}({\bf \hat D}){\bf w}_\textsf{y}({\bf \hat
 D}), \qquad  {\hat {\bf U}}_k\equiv {\hat {\bf U}}({\bf y}_{k}).
 \label{eq:updateC}
\end{eqnarray}
\noindent Here,  use was made of the definitions
\begin{eqnarray*}
 \label{eq:defs}
 &&\hspace{-0.5in}{\bf w}_\textsf{y}({\bf \hat D})  =
    {\bf \hat \Sigma}_{\textrm{pr}}^{-1} \, {\bf c}_{\textrm{pr}}
    + h\sum_{k = 0}^{K - 1}\left[ {\bf \hat U}_{k}^T \, {\bf \hat D}^{-1} \, \dot{{\bf
 y}}_{k}-  \frac{1}{2}  {\bf v}({\bf y}_k)\right], \\
 &&\hspace{-0.5in}{\bf \hat  \Xi}_\textsf{y}({\bf \hat  D})
    =  {\bf \hat  \Sigma}^{-1}_{\textrm{pr}} + h \, \sum_{k = 0}^{K - 1} {\bf \hat U}_{k}^{T} \, {\bf \hat D}^{-1} \,
    {\bf \hat U}_{k}.
    \nonumber
\end{eqnarray*}
We repeat  this two-step optimization procedure iteratively,
starting from arbitrary prior values ${\bf c}_{\textrm{pr}}$ and
${\bf \hat \Sigma}_{\textrm{pr}}$. We emphasize that a number of
important parameters of the decomposition of the original signal
(e.g.\ the bandwidth, order of the filters and scaling parameters
$a_{ki}$) have to be selected to provide the best fit to the
measured time series $\{ {\bf s}(t_k)\}$.
\begin{figure} 
\includegraphics[width=8cm,height=6cm]{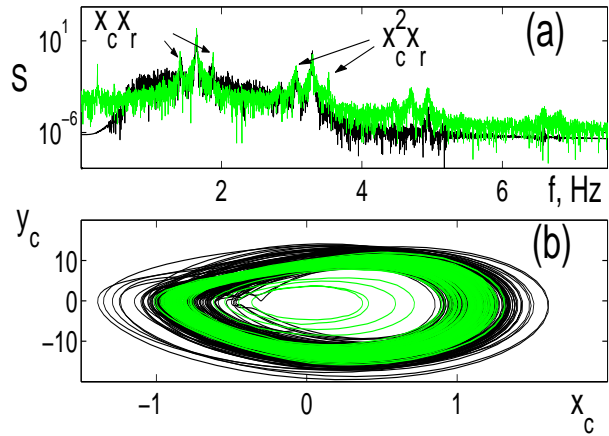}
\caption{\label{fig:cardiac} (a) Power spectra of cardiac
oscillations obtained from measured data (black line) and from the
synthesized model signal (green line). Arrows summarize
combinational frequencies recovered in our analysis, corresponding
to the nonlinear cardio-respiratory interaction. (b) Limit cycles
of the cardiac oscillations $(x_c(n),y_c(n)$ obtained from
measured data (black line) and the synthesized signal (green
line).}
\end{figure}
The parameters of the model (\ref{eq:cardio-respiratory}) can now
be inferred directly from the measured time series of blood
pressure, yielding the values shown in the first row of Table
\ref{tab:1}. The spectra of the inferred, $x_r(t)$, and the
measured, $s_r(t)$, cardiac oscillations are compared in Fig.
\ref{fig:cardiac}. Similar results are obtained for the
respiratory oscillations. In particular, the parameters of the
nonlinear coupling and of the noise intensity of the cardiac
oscillations are $\beta_{20}=2.2, \beta_{21}=0.27$,
$\beta_{22}=-8.67$, and $D_{22}=8.13$ ; here we use a
double-indexing scheme for the coefficients of the linear
expansion (\ref{eq:basis}), the scheme being evident from the
caption in Table \ref{tab:1}. It is clear that there is a close
resemblance between the peaks at the basic and combinational
frequencies, $n f_c+ m f_r$, in the power-spectra. A similarly
close resemblance is found for respiratory oscillations, $s_r(t)$
and $x_r(t)$, respectively (not shown).

The frequency content can be reproduced from a univariate signal
$s(t)$ because for $f_r\ll f_c$ it can be written in the form:
$s(t)\approx s_r(t)+A_c(t)\cos(f_c t+\theta_c(t))+\ldots$, here
$A_c(t),\theta_c(t)$ are slow amplitude and phase and the omitted
terms oscillate at multiples of $f_c$. Fast-oscillating terms in
this expansion  correspond to a cardiac signal $s_c(t)$ and this
ensures the validity of the signal decomposition
$s(t)=s_r(t)+s_c(t)$, with components corresponding to weakly
coupled nonlinear oscillators.

\begin{table}[ht]
   \begin{tabular}{|c|c|c|c|c|c|c|c|}\hline
   $\alpha_{20}$ & $\beta_{20}$ & $\alpha_{21}$ & $\beta_{21}$ & $\alpha_{22}$ & $\beta_{22}$ & $D_{11}$  & $D_{22}$\\
        \hline
   0.12  &  2.20  &  0.048  &  0.27 &  -0.066 &  -8.67  &  0.18  &  8.13\\
        \hline
   0.12  &  2.41  &  0.048  &  0.28 &  -0.070 &  -8.61  &  0.18  &  8.14\\
        \hline
    2.9\%  &  9.3\%  &  1.8\%  &  5.6\% &  5.2\% &  0.7\%  &   0.2\%  &  0.2\% \\
        \hline
        \end{tabular}
    \caption{Coefficients
    corresponding to the last three base functions in
    (\ref{eq:basis}),
    $\{x_rx_c,x_r^2x_c,x_rx_c^2\}$, with $\{\alpha_i\}$ corresponding to the respiration
    coupling to cardiac rhythm and $\{\beta_i\}$ to the
    cardiac oscillation coupling to respiration. The top row gives coefficients
    inferred from measured data. The middle row represents coefficients
    inferred from synthesized data, obtained as an average of 100 non-overlapped 1600~s blocks.
    Each block includes 160000 points with a sampling time 0.01 sec. The estimation error is shown
    in the bottom line.}
    \label{tab:1}
\end{table}

To validate these results we consider a synthesized signal
$x(t)=x_r(t)+x_c(t)$ where $x_r(t)$, $x_c(t)$ are obtained using
numerical simulations of the model (\ref{eq:2vdp_1}) with the
parameters taken from the inference. We now repeat the full
inference procedure to estimate nonlinear coupling parameters in
(\ref{eq:2vdp_1}) by using the synthesized univariate signal
$x(t)$ as a time-series data input $s(t)$. This gives us the
following estimates for the parameters of cardiac oscillations
$\beta_{20}=6.32, \beta_{21}=0.49$, $\beta_{22}=6.03$, and
$D_{22}=3.44$, which differ from the values in the first row of
Table \ref{tab:1}, but provides a correct estimation of the order
of magnitude of the absolute values of the measured parameters.
The main source of error here is the fact that we have to
reconstruct the state of multidimensional system using the
univariate signal.

If the state of the system was known the accuracy of inference
could be arbitrary high~\cite{Smelyanskiy:submitted}. To
illustrate this point we use the synthesized time-series
$\{x_r(t), x_c(t), y_r(t), y_c(t)\}$ as bivariate data for two
coupled oscillators to infer parameters of the model
(\ref{eq:2vdp_1}). The results are summarized in the second row of
Table \ref{tab:1}. It can be seen that the values of the
parameters can be estimated with relative error of less than 10\%.
In particular, the relative error of estimation of the noise
intensity is now below 4\%. The accuracy of the estimation can be
further improved by increasing the total time of observation of
the system dynamics. The decomposition problem could of course be
eliminated by using bivariate cardiovascular data, which are now
commonly available.

The relative magnitudes of the parameters obtained,
$|\beta_{i}|>|\alpha_{i}|$, indicate that respiration influences
cardiac activity more strongly than vice versa, consistent with
the results of methods specifically developed for detecting the
coupling directionality of interacting oscillators
\cite{Rosenblum:02Palus:03a}, and with direct physiological
observations. Furthermore, the presence of non-zero quadratic
terms is consistent with recent results obtained by time-phase
bispectral analysis \cite{Jamsek:03Jamsek:04}. The frequency and
amplitude variability of the main oscillatory components
\cite{Stefanovska:99a} is implicitly captured within the coupling
terms and noise. We find that the present model class is able to
reproduce, not only the coupling directionality, but also to a
large extent the 1:7 and 1:8 cardio-respiratory synchronization
properties of the measured data, as will be discussed in detail
elsewhere.

We would like to mention that reported method is only a first step
in the direction of developing path-integral based approach to the
dynamical inference of stochastic nonlinear models. It was
verified on a number of model systems and has demonstrated stable
and reliable inference of a broad class of models with high
accuracy (see e.g. \cite{Smelyanskiy:submitted}). However, the
method in its present form has a number of limitations. For
example, to include frequencies lower then the frequency of
respiration as well as to account for feedback mechanism of
control from the nervous system will require for an extension of
the model class used in the paper. In particular, it will require
to include new degrees of freedom, time-delay functions and
non-polynomial basis functions, possibly a non-white noise and
non-parametric model inference. However, the technique can be
readily extended to encompass mentioned above situations.

In summary, we have solved a long-standing problem in physiology:
inference of a nonlinear model of cardio-respiratory interactions
in the presence of fluctuations. Our technique estimates
simultaneously the strength and directionality of coupling, and
the noise intensity in the cardio-respiratory interaction,
directly from measured time series. It can in principle also be
applied to any physiological signal. Our solution is facilitated
by an analytic derivation of the likelihood function that
optimally compensates noise-induced errors in continuous dynamical
systems. It has enabled us to effect the first application of
nonlinear stochastic inference to identify a dynamical model from
real data.

This work was supported by NASA CICT IS IDU project (USA), by the
Leverhulme Trust and by EPSRC (UK), by the M\v SZ\v S (Slovenia),
and by INTAS.



\begin{thebibliography}{10}
\bibitem{Decomposition}
Alternative approaches could include e.g.\ empirical mode
decomposition, Karhunen-Lev\'e decomposition, or independent
component analysis.
\end{thebibliography}


\end{document}